\documentclass[%
 reprint,
superscriptaddress,
nofootinbib,
 amsmath,amssymb,
 aps,
pra,
]{revtex4-1}
\usepackage{refcount}
\usepackage{graphicx}% Include figure files
\usepackage{dcolumn}% Align table columns on decimal point
\usepackage{bm}% bold math
\usepackage{color}
\usepackage{mathrsfs}
\usepackage[english]{babel}
\usepackage{amsthm}
\usepackage{dsfont}
\usepackage{float}
\usepackage{textcase}
\usepackage{titlecaps}
\usepackage{balance}
\usepackage{flushend}
\usepackage{txfonts}

\usepackage[dvipsnames]{xcolor}
\usepackage[hidelinks,colorlinks=true, linkcolor=WildStrawberry, urlcolor=WildStrawberry, citecolor=Emerald, linktocpage=true, breaklinks=true, bookmarksopen=true]{hyperref}

\newcommand{\BA}[1]{\textcolor{purple}{ #1}}

\addto\captionsenglish{}

\usepackage{amsmath}
\begin{document}

\begin{abstract}
Quantum metrology exploits quantum mechanical effects to increase the precision of measurements of physical quantities. A wide variety of applications are currently being developed for scientific and technological purposes, however, most research relies on the use of highly entangled resource states that are challenging to generate and control in a given physical system. Here, we study the use of weighted graph states as more accessible resources for quantum metrology, which yield a favorable precision beyond the classical limit, approaching the Heisenberg limit. We find a notable robustness to variation in weights and less challenging weight requirements compared to standard graph states, which require a maximal weight at all edges. Both of these aspects reduce the practical demands in a physical setup, with the latter implying significantly less entanglement is required to gain a quantum advantage in metrology. We study the quantum Fisher information and optimized estimator variance of two identified sub classes of weighted graph states for an arbitrary number of $N$ qubits, providing analytical forms and investigating their scaling. Our work opens up opportunities for using weakly entangled states in quantum-enhanced metrology.       
\end{abstract}

\preprint{APS/123-QED}

\title{Weighted graph states as a resource for quantum metrology}% Force line breaks with \\

\author{B. J. Alexander}
 \email{alexander.byronj@gmail.com}%Lines break automatically or can be forced with \\
 \affiliation{%
 Department of Physics,
 Stellenbosch University, Matieland 7602, South Africa 
}
\author{\c{S}. K. \"{O}zdemir}
% \email{john.bollinger@nist.gov}
\affiliation{Department of Electrical and Computer Engineering, Saint Louis University, St.~Louis, Mo 63103, USA}

\author{M. S. Tame}%
% \email{mstame@sun.ac.za}
\affiliation{%
 Department of Physics,
 Stellenbosch University, Matieland 7602, South Africa 
}

\date{\today}% It is always \today, today,
             %  but any date may be explicitly specified
\maketitle

%%%%%%%%%%%%%%%%%%%%%%%%%%%%
%%%%%%%%%%%%%%%%%%%%%%%%%%%%
%%%%%%%%%%%%%%%%%%%%%%%%%%%%
%%%%%%%%%%%%%%%%%%%%%%%%%%%%
\section{Introduction}
Quantum metrology is primarily concerned with understanding how quantum mechanical effects, such as entanglement \cite{horodecki2009quantum} and spin-squeezing \cite{kitagawa1993squeezed, ma2011quantum}, can be controlled and utilized for the purpose of reaching a precision in measurement of a physical parameter beyond classical limits \cite{horodecki2009quantum, toth2014quantum, giovannetti2006quantum, degen2017quantum}. It is therefore important to identify and characterize classes of entangled states that can provide a quantum advantage for well-defined metrological tasks. Recently, the class of graph states~\cite{hein2006entanglement, briegel2009measurement} has been shown to be particularly well-suited for applications in quantum metrology \cite{shettell2020graph, friis2017flexible}. This is practically relevant, as graph states have been physically realized in a wide range of physical settings, including in ion-traps \cite{lanyon2013measurement,pogorelov2021compact,moses2023race,ringbauer2025verifiable,kang2024entanglement}, neutral atoms \cite{mandel2003controlled, cooper2024graph}, nitrogen-vacancy centers \cite{cramer2016repeated}, quantum dots~\cite{Cogan22, istrati2020sequential, huet2025deterministic}, single atoms~\cite{Thomas22, thomas2024fusion}, superconducting qubits \cite{song201710, mooney2019hill,ferreira2024deterministic} and photonic systems \cite{wang201818}. However, there is a high level of control needed in these experiments to obtain good quality resource states and there is a key question of whether resources that are less challenging to produce can provide comparable functionality.

In this work we address this important question by studying weighted graph states \cite{hein2006entanglement, hartmann2007weighted, dur2005entanglement}, which are characterized by a weak pairwise entanglement between qubits at vertices, realized via controlled phase gates with arbitrary phase - the value of the phase representing the weight. It is known that while not as entangled as standard graph states (with full weights), weighted graph states can be used for universal and deterministic quantum computing \cite{gross2007measurement,kissinger2019universal}, and therefore, although somewhat indirectly, they have already been shown to be viable resources for quantum metrology, since from larger weighted graph state structures, one can produce fully-weighted graph states, which can then be used according to Refs.~\cite{shettell2020graph, friis2017flexible}. In our work, however, we are interested in using weighted graph states as a resource for metrology directly, as opposed to an intermediary resource for producing fully-weighted graph states. Our work goes beyond that in Ref.~\cite{xue2012spin} by using recent advances
in analyzing and quantifying the sensing performance of quantum states~\cite{toth2014quantum, shettell2020graph,friis2017flexible}. In particular, we look at the quantum Fisher information (QFI) to quantify the sensing advantage of weighted graph states, which is more rigorous from a metrological standpoint. In doing so, we reduce the practical requirements of quantum metrology applications which conventionally utilize standard graph state resources. This is in the spirit of other recent proposals using weighted graph states, including for randomizing inputs \cite{plato2008random}, which is utilized extensively in quantum information -- from modelling noisy quantum systems \cite{emerson2005scalable} to information leakage of black holes \cite{hayden2007black}. Further examples of weighted graph state applications include utilization as a resource state for measurement-based Toffoli gates~\cite{tame2009compact}, for generating perfect maximally entangled photonic Greenberger-Horne-Zeilinger (GHZ) states~\cite{frantzeskakis2022extracting}, as a variational set for finding ground states of spin systems~\cite{Anders07}, generating novel entanglement measures~\cite{Hajdusek13,Hajdusek14}, studying sub-universal instantaneous quantum polynomial time models for determining quantum computational advantages~\cite{hayashi2019verifying}, and uncovering transitions in variable-range interacting models \cite{ghosh2024entanglement}.

In our study, we consider the class of weighted graph states as an entangled resource for the metrological task of phase estimation \cite{toth2014quantum}, where the estimated parameter $\theta$ is the encoded phase associated with a state having undergone some unitary evolution, that is, \begin{align}
    \rho \mapsto \rho_{\theta} := \hat{U}(\theta) \rho \hat{U}^\dagger(\theta).
    \label{A}
\end{align}
Essentially, to estimate the parameter $\theta$ to highest precision, we seek to identify and characterize metrologically relevant sub-classes of weighted graph states which consistently demonstrate a quantum advantage in achieving a precision, or resolution, below the classical limit, also known as the standard quantum limit (SQL). These sub-classes are uniquely characterized by their graph geometry. In doing so, we generalize the main results of a recent study that focused on utilizing fully-weighted graph states for quantum sensing \cite{shettell2020graph}. We show that the physical requirements can be significantly relaxed in the sense that the weighted graphs need only a fraction of the entanglement of their fully weighted counterparts. Our work also goes beyond Ref.~\cite{xue2012spin}, as we provide the optimal
measurements that can be performed to achieve a given level of enhancement
in sensing, which is of practical importance and experimental relevance. 
\begin{figure*}[tpb]
\hspace*{-1.25cm}
  \centering
  \begin{tabular}{ccc}
    \includegraphics[width=1\textwidth]{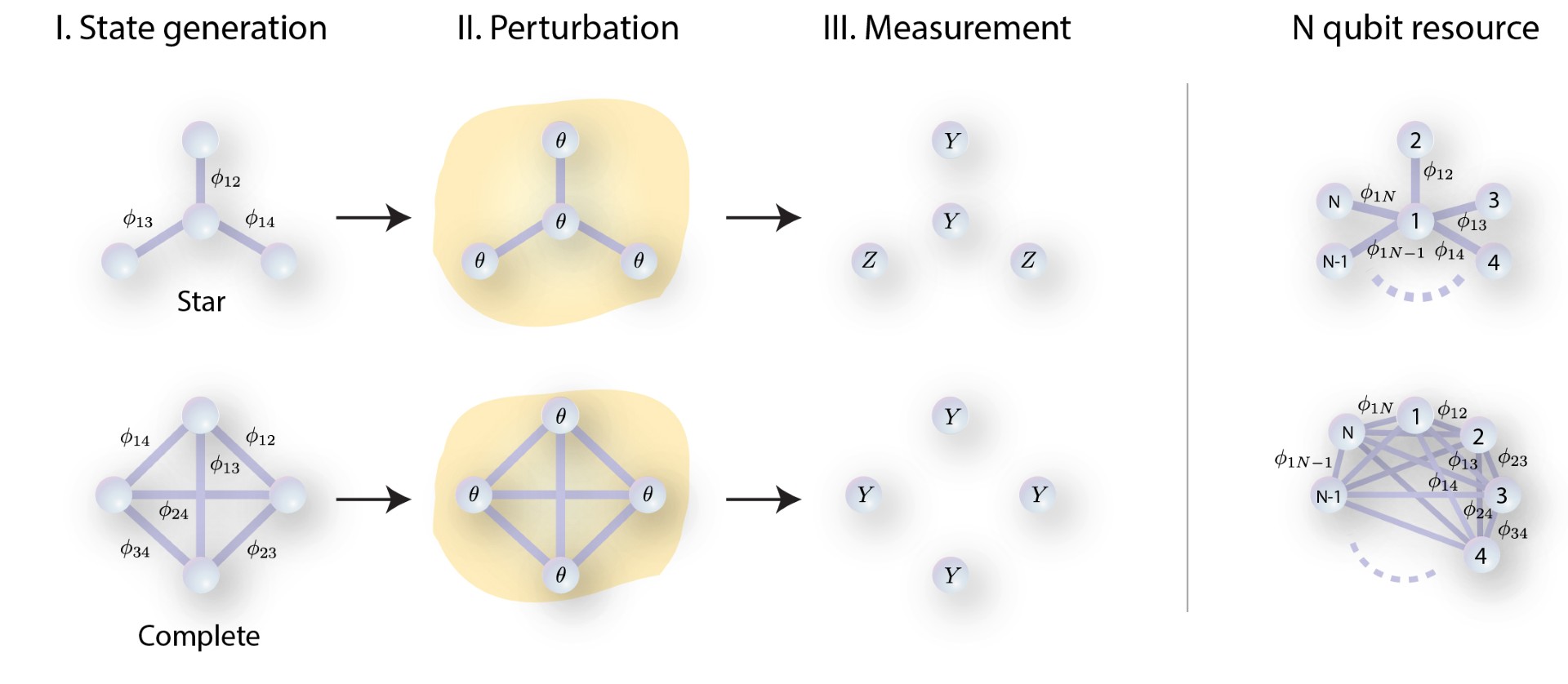}
    \label{AA}
  \end{tabular}
  \caption{The main steps of the phase estimation scheme which utilize the example star and complete weighted graph state sub-classes (4-qubit class representatives shown here): (I) State generation - the vertices correspond to qubits initialized to $|+\rangle$ and the $\phi_{ab}$-edges  connecting the vertices represent controlled phase-$\phi$ gates. Note that for the analytical study in this section, we assume uniform edge weights, i.e., $\phi_{ab} = \phi$ for all edges, whereas in the subsequent numerical analysis we allow randomly chosen edge weights $\phi_{ab}$; (II) Perturbation - encode the phase $\theta$ by a unitary transformation; (III) Measurement - execute the optimal (Pauli) measurement operators which minimizes the uncertainty, that is the variance $(\Delta \theta)^2$ in estimating $\theta$. More specifically, the optimal measurement operators are found to be $\hat{Y} \otimes \hat{Z} \otimes \hat{Y} \otimes \hat{Z}$ and $\hat{Y} \otimes \hat{Y} \otimes \hat{Y} \otimes \hat{Y}$, for the $N=4$ star and complete weighted graph states, respectively. The right hand side depicts the generalization to $N$ qubits for the two sub-classes. The two sub-classes are locally equivalent only when all edge weights are maximal (see the main text for details).}
  \label{1}
\end{figure*}

Furthermore, we conduct a numerical robustness study, where the interaction weightings between qubits (hereafter denoted by $\phi_{ab}$) are independent and identically distributed (IID), that is $\phi_{ab}\sim\mathcal{N}\big(\mu, \sigma\big)$ (denoting a Gaussian distribution with mean $\mu$ and standard deviation $\sigma$). We thereby demonstrate that a quantum advantage is maintained for weightings within an interval centered on the identified optimal uniform weighting; thus even an imperfect preparation of the identified sub-classes with a reduced entanglement still consistently goes below the classical SQL. 

In section II, we provide an outline of phase estimation, a brief overview of some requisite concepts in quantum metrology, as well as the basic theory underpinning the numerical models employed. In section III, we discuss the numerical results which motivate the classifications of various relevant weighted graph states. In section IV, we provide the associated analytic forms of the QFI and optimized estimator variance. In section V, we conclude with a summary of our results and an outlook on future work.

%%%%%%%%%%%%%%%%%%%%%%%%%%%%
%%%%%%%%%%%%%%%%%%%%%%%%%%%%
%%%%%%%%%%%%%%%%%%%%%%%%%%%%
%%%%%%%%%%%%%%%%%%%%%%%%%%%%
\section{Theory}  
To address the quantum metrology phase estimation task, we consider a scalar parameter $\theta$, encoded in a multi-qubit state $\rho$, by a unitary transformation $\hat{U}$ (see Eq.~(\ref{A})) which is characterized by the Hamiltonian operator $\hat{H}$, that is, \begin{align}
    \hat{U}(\theta):=\text{exp}\big(-i \hat{H} \theta/ \hbar\big).
    \label{B}
\end{align}
By measuring the expectation value of some Hermitian operator $\hat{A}$, the precision of estimating the unitary phase parameter $\theta$, for a single-shot measurement, is quantified by the estimator variance $(\Delta \theta)^2$ for an unbiased estimator $\hat{\theta}$ (see Appendix A for more details on point estimation) via the error-propagation formula \cite{toth2014quantum}
\begin{align}
    (\Delta\theta)^2:=\frac{(\Delta A)^2}{\big|\partial_{\theta} \langle \hat{A} \rangle \big|^2},
    \label{D}
\end{align}
where $\langle \hat{A} \rangle$ and  $(\Delta A)^2:=\langle \hat{A}^2 \rangle-\langle \hat{A} \rangle^2$ denote the operator expectation value and variance, respectively. It is clear from Eq.~(\ref{D}) that the estimation precision depends on the variance $(\Delta A)^2$ and the sensitivity of $\langle \hat{A} \rangle$ to changes in $\theta$. Note that for an $N$-qubit state, the lower bound of the estimator variance $(\Delta \theta)^2$ for separable resource states, scales as $\sim\frac{1}{N}$ (the SQL), whilst the lower bound for entangled resource states scales as $\sim\frac{1}{N^2}$ (the Heisenberg limit (HL))~\cite{shettell2020graph}.

In this study, we restrict the Hermitian measurement operators to the class of
tensor products of single-qubit Pauli operators. For each qubit \(i\), the local
operator is chosen from
\(\hat{A}_{i} \in \big\{ \hat{\mathds{1}}, \hat{X}, \hat{Y}, \hat{Z} \big\}\).
To that end, for an \(N\)-qubit system, the measurement observables are taken as
\(2^N \times 2^N\) tensor-product operators,
\begin{equation}
    \hat{A} := \hat{A}_{1}\otimes\hat{A}_{2}\otimes \cdots \otimes\hat{A}_{N}.
    \label{OpA}
\end{equation}
This type of local measurement is easier to perform in a practical setting and in line with measurement-based techniques~\cite{hein2006entanglement}.

The class of weighted graph states are chosen as the entangled resource to address the phase estimation task. Let $G=(V,E)$ be a graph, where $V$ denotes the set of vertices (qubits in this study) and $E\subseteq\big\{ \{a,b\}|~a,b \in V;~a \neq b\big\}$ the set of edges (controlled phase-$\phi$ gates in this study). The weighted graph state $|G\rangle$, with pairwise interaction weightings $\phi_{ab}$, associated with graph $G$, is defined by \begin{equation}
    |G\rangle := \prod_{\{a,b\}\in E}\hat{U}_{ab}(\phi_{ab})|+\rangle^{\otimes V},
    \label{E}
\end{equation} 
where $|+\rangle :=\frac{1}{\sqrt{2}}(|0\rangle  +|1\rangle)$ and
\begin{equation}
    \hat{U}_{ab}(\phi_{ab}):= \text{exp}\big(i\phi_{ab}\hat{H}_{ab}/\hbar\big). 
    \label{F}
\end{equation}
Throughout, the edge weights $\phi_{ab} \in [0, 2\pi)$ are phase angles (in radians), defined modulo $2\pi$. 
The Hamiltonian which generates the unitary operator in Eq.~(\ref{F}) is 
\begin{equation}
\hat{H}_{ab} := |1\rangle_{a}\langle 1 | \otimes |1\rangle_{b}\langle 1|,
\label{HGS}
\end{equation}
which gives $\hat{U}_{ab}(\phi_{ab})=\text{diag}(1,1,1,e^{i \phi_{ab}})$. Physical systems that implement such an arbitrary controlled-phase shift include photonic \cite{young2011quantum,firstenberg2013attractive,firstenberg2016nonlinear,tiarks2016optical,thompson2017symmetry,tiarks2019photon,sagona2020conditional}, superconducting \cite{o2025deterministic} and atomic systems \cite{qin2025scaling}, where lower values of $\phi_{ab}$ may be easier to realize, as they require shorter interaction times and are therefore less affected by noise.
Note that by definition, the class of fully-weighted graph states assumes that $\phi_{ab}=\pi$ for all $\{a,b\}\in E$ (see Ref.~\cite{hein2006entanglement}). Fig.~\ref{A} presents a schematic outline of the utilization of two example weighted graph state sub-classes (with $N=4$) for phase estimation, namely the star and complete graphs, together with their generalization to $N$ qubits. The $N$-qubit star weighted graph state is characterized by a central qubit connected to each of the remaining $N-1$ qubits, while the $N$-qubit complete weighted graph state is characterized by connecting all qubits to each other.

For the purpose of quantifying the information resources of relevant quantum states used for phase estimation, we study the quantum Fisher information (QFI), which is the quantum analogue \cite{petz2011introduction, paris2009quantum} of the classical Fisher information \cite{helstrom1969quantum, braunstein1994statistical}, an important quantity in quantum metrology \cite{lu2015robust, toth2014quantum}. Consider some unitary transformation as given by Eq.~(\ref{B}), the QFI with respect to some Hamiltonian operator $\hat{H}$, for some state $\rho$, is given by \cite{helstrom1969quantum, braunstein1994statistical, paris2009quantum}
\begin{align}
    \mathcal{Q}(\rho, \hat{H}):= 2\sum_{\substack{j,k=1 \\ \lambda_{j}+\lambda_{k} \neq 0}}\frac{(\lambda_j-\lambda_k)^2}{\lambda_j+\lambda_k}\big|\langle j | \hat{H}|k\rangle\big|^2,
    \label{QFI}
\end{align}
where $\lambda_i$ and $|i\rangle$ are respectively the eigenvalues and eigenvectors of $\rho$. It can also be shown that 
\begin{align}
    \mathcal{Q}(\rho, \hat{H}) \leq 4(\Delta H)^2
    \label{H},
\end{align}
where equality holds for pure states \cite{paris2009quantum}. Moreover, the QFI serves to constrain the achievable precision, by setting a lower-bound on the estimator variance given by Eq.~(\ref{D}). The phase estimation precision, for any single-shot measurement, is restricted by the well-known quantum Cram\'{e}r-Rao bound~\cite{helstrom1969quantum, braunstein1994statistical, paris2009quantum}, that is, 
\begin{align}
    (\Delta \theta)^2 \geq \frac{1}{\mathcal{Q}(\rho, \hat{H})}.
     \label{CR}
\end{align}

In this study, we consider the Hamiltonian operator 
\begin{align}
    \hat{H}:=\frac{1}{2}\sum_{j=1}^{N}\hat{X}_{j},
    \label{C}
\end{align}
where $\hat{X}_{j}$ is the Pauli-$X$ operator acting on the $j$-th qubit. In the sections that follow, we investigate the impact of a unitary rotation on a collection of qubits described by the Hamiltonian given by Eq.~(\ref{C}). To maximize the effect of the unitary phase $\theta$ on the state such that its influence is more detectable, we assume that all qubits undergo the same Pauli-$X$ rotation of magnitude~$\theta$. Although it is possible to apply the rotation to only subsets of qubits, we have chosen to defer this more general case to a future study. In our analysis we choose the encoding Hamiltonian given in Eq.~(\ref{C}) rather than the locally optimal generator at $\phi = \pi$, which is $\frac{1}{2}(\hat{Z}_{1} + \sum_{j=2}^{N}\hat{X}_{j})$. This choice is motivated by the study of graph states for metrology in Ref.~\cite{shettell2020graph}, where a simple, uniformly applied Hamiltonian is adopted so that each qubit experiences the same local evolution. Using Eq.~(\ref{C}) allows us to remain in this standard setting, facilitates direct comparison with that work, and highlights the metrological performance attainable with experimentally realistic, homogeneous encodings, without tailoring the generator to the specific graph structure. It is important to note, however, that at $\phi = \pi$ the corresponding optimal generator yields a QFI of $N^{2}$, whereas our choice is not strictly optimal for finite $N$ (an analytic form of the QFI is provided later). Nevertheless, as $N \to \infty$ the QFI of the non-optimal encoding becomes asymptotically equivalent to the optimal one up to a vanishing relative error. 

For pure states, using Eq.~(\ref{H}), and assuming the Hamiltonian operator Eq. (\ref{C}), the QFI reads as \cite{shettell2020graph}
\begin{align}
     \mathcal{Q}(\rho,\hat{H})= \sum_{j,k=1}^{N}\big[ \text{Tr}(\hat{X}_{j}\hat{X}_{k}\rho)-\text{Tr}(\hat{X}_{j}\rho)\text{Tr}(\hat{X}_{k}\rho)\big].
     \label{QFI2}
\end{align}

For separable states, we have the QFI upper bound
\begin{equation}
\mathcal{Q}(\rho, \hat{H})\leq N.
\label{sep}
\end{equation}
States which violate Eq.~(\ref{sep}) are entangled  (see Ref.~\cite{toth2014quantum}). This defines the SQL in terms of the QFI. The Cram\'{e}r-Rao bound in Eq.~(\ref{CR}) then yields the SQL scaling in terms of the estimator variance as mentioned previously, that is, $(\Delta \theta)^2 \sim \frac{1}{N}$.

\begin{figure}[tb]
\hspace*{0cm}
  \centering
  \begin{tabular}{ccc}
   \includegraphics[width=79mm,scale=0.99]{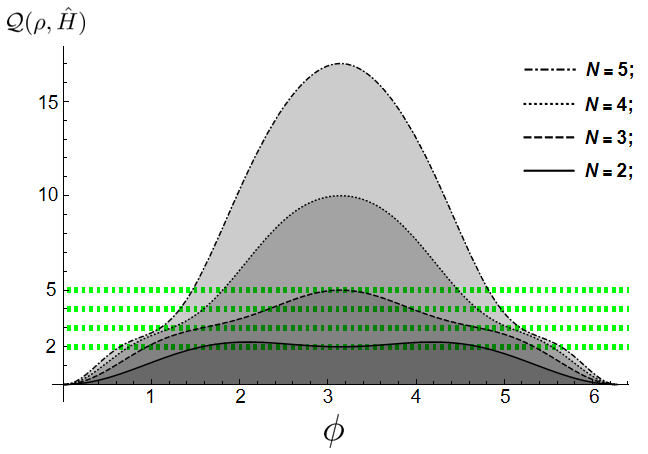}
  \end{tabular}
   \caption{The QFI of weighted star graph states, with $\hat{H}$ given by Eq.~(\ref{C}). The dashed horizontals denote the corresponding QFI upper bounds for separable states (SQL) of size $N \in \{2,3,4,5\}$.}
  \label{Star_QFI2}
\end{figure}

\begin{figure}[t]
\hspace*{0cm}
  \centering
  \begin{tabular}{ccc}
  \includegraphics[width=79mm,scale=0.99]{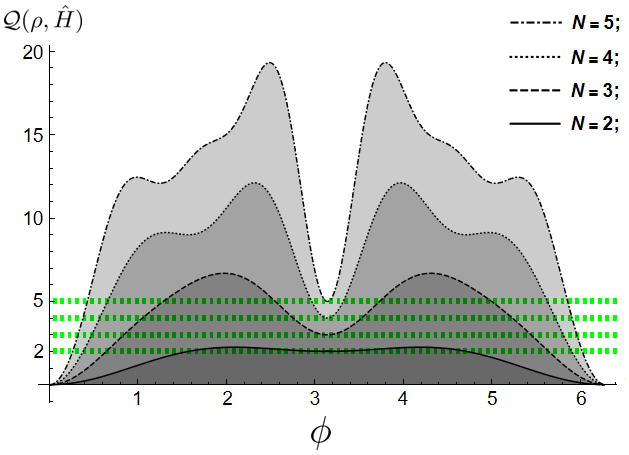}
  \end{tabular}
  \caption{The QFI of weighted complete graph states, with $\hat{H}$ given by Eq.~(\ref{C}). The dashed horizontals denote the corresponding QFI upper bounds for separable states (SQL) of size $N \in \{2,3,4,5\}$.}
  \label{Comp_QFI2}
\end{figure}

Fig.~\ref{Star_QFI2} presents an initial study demonstrating that the QFI of the class of star weighted graph states is robust when the uniform weighting $\phi$ is chosen around the standard weighting of $\pi$.  Fig.~\ref{Comp_QFI2} demonstrates that the QFI of complete weighted graph states with a reduced uniform weighting $\phi \in [0,\pi )$, is often greater than that of the standard complete graph state which assumes $\phi = \pi$. Furthermore, by restricting to local Pauli measurements, we find that one can gain an advantage, as highlighted by the QFI for $N \geq 3$ in Fig.~\ref{Star_QFI2} and Fig.~\ref{Comp_QFI2}, which we examine in more detail in the subsequent section by evaluating the precision $(\Delta \theta)^2$ (defined in Eq.~(\ref{D})).  The analytic forms of the QFI for the $N$-qubit star and complete weighted graph states, assuming Hamiltonian Eq.~(\ref{C}), are presented later in section IV.
\begin{figure*}[tb]
\hspace*{-0.375cm}
  \centering
  \begin{tabular}{ccc}
    \includegraphics[width=1.04\textwidth]{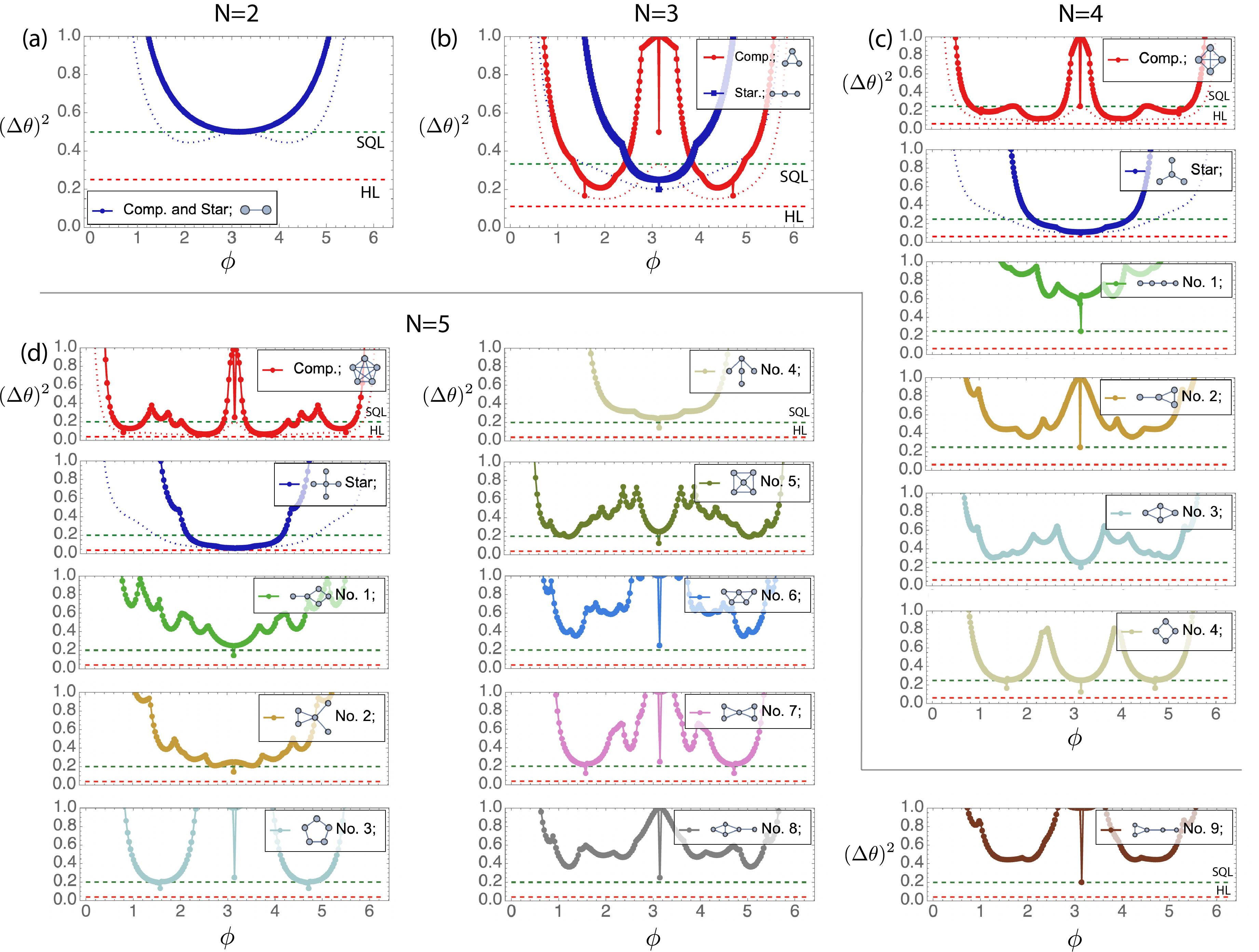}
  \end{tabular}
  \caption{The optimal estimator variance $(\Delta \theta)^2$, restricting to Pauli measurements, over the set of connected uniformly weighted graph states which go below the SQL (assuming $\theta =0.001$ and system size $N \in \{2,3,4,5\}$). It is important to note that small $\theta$ is the most interesting regime for quantum metrology and that any $\theta$ can be shifted to this point in principle via an offset phase. Dashed horizontal green and red lines indicate the SQL and HL, respectively, while the dotted curves show the Cram\'{e}r-Rao bounds computed from the analytical QFI expressions derived for star and complete weighted graph states in Section IV. The points of discontinuity are discussed in the main text.}
  \label{B1}
\end{figure*}
In the following section, we conduct a numerical analysis for varied system size $N$, with the purpose of identifying, based on their graph geometry, the classes of weighted-graph states which consistently go below the classical SQL in terms of the precision $(\Delta \theta)^2$. Essentially, we seek to minimize Eq.~(\ref{D}). In doing so, we will motivate the selection of the star and complete weighted graph state sub-classes as being optimal for phase estimation. 

%%%%%%%%%%%%%%%%%%%%%%%%%%%%
%%%%%%%%%%%%%%%%%%%%%%%%%%%%
%%%%%%%%%%%%%%%%%%%%%%%%%%%%
%%%%%%%%%%%%%%%%%%%%%%%%%%%%
\section{Numerical Analysis}
For an $N$-qubit uniformly weighted graph state, where $\phi_{ab} = \phi \in [0, 2\pi)$ for all $\{a,b\}\in E$, the numerical optimization protocol traverses all combinations of Pauli measurement operators given by Eq.~(\ref{OpA}), with the purpose of identifying the optimal operator, characterizing relevant graph state sub-classes and corresponding weighting intervals, which yield local minimum estimator variance values (using Eq.~(\ref{D})) below the SQL. 

The results presented in Fig.~\ref{B1} consider only graph geometries which go below the SQL for some \BA{$\phi \in [0, 2\pi)$}. For a given graph state, the optimal Pauli measurement is found to vary as a function of $\phi$ (as we seek to minimize Eq.~(\ref{D})). We consider $N=2,~3,~4$ and $5$ to begin with. From these results for low $N$, we identify two classes of weighted graph states, characterized  by their distinct graph geometries (as a function of $N$) which produce favorable precision results. In Fig.~\ref{B1} the sub-class representatives are labelled `Comp.' and `Star', denoting elements of the complete and star weighted graph state classes, respectively.  

It is important to note the equivalence of some of the graphs shown in Fig.~\ref{B1}. In graph theory, an adjacent vertex of a given vertex $v \in V$ is a vertex that is connected to $v$ by an edge. On the other hand, the neighborhood of $v$ is the set of all vertices that are adjacent to $v$; the subgraph induced by this set is called the neighborhood subgraph of $v$ (see Ref.~\cite{west2001introduction, wilson1979introduction}). In the context of graph states, which are characterized by uniform $\pi$ weights, local complementation modifies the connections among the neighbors of a chosen vertex: pairs of neighbors that were previously unconnected become connected, while pairs that were previously connected have their edges removed. By changing the set of edges $E$, this operation changes the connectivity of the graph, while preserving the graph state up to local operations. Clifford operators are a particular subset of local unitary operators that are widely used in quantum information theory. They are defined as unitary operators that map stabilizer states to stabilizer states, where a stabilizer state is a quantum state that is invariant under a group of commuting operators known as the stabilizer group. Stabilizer states play a central role in the theory of graph states (see Ref.~\cite{hein2006entanglement}). More formally, graph states $|G\rangle$ and $|G’\rangle$ are local Clifford (LC) - equivalent if they are related by some local Clifford unitary $\hat{U}$ such that $|G’\rangle = \hat{U}|G\rangle$. The action of a local Clifford operation on a graph state can be equivalently described in terms of some local complementation of the graph state. It follows that two graph states are LC-equivalent if and only if the corresponding graphs are related by a sequence of local complementations \cite{hein2006entanglement}.  

Some of the weighted graph states shown in Fig.~\ref{B1}, when restricting to uniform $\phi = \pi$ weightings (which define standard graph states), are known to be local unitary (LU) equivalent through local complementation (a series of LC operations), or more generally, local operations and classical communication (LOCC) (see Ref.~\cite{nielsen2002quantum}), but when considering them as weighted graph states these graphs are generally locally inequivalent and must be treated independently. A shared property of the fully-weighted (with $\phi=\pi$) star and complete graph states, is that they are both LU equivalent to the well-known maximally entangled Greenberger-Horne-Zeilinger (GHZ) state. Applying Hadamard unitary operators to all but one qubit of the GHZ state yields the star graph state. There exists a local Clifford unitary operator, such that when applied to the star graph state, yields the complete graph state~\cite{hein2006entanglement}.  

By increasing the system size $N$ of star and complete weighted graph states (see right hand side of Fig.~\ref{1}), it becomes apparent from Fig.~\ref{B1} that the optimal measurement operators for the aforementioned sub-classes follow a predictable pattern as a function of $N$. Figure~\ref{B1} also shows that certain weighting intervals yield superior precision results depending on the graph class. 
For both sub-classes (complete and star), the optimal measurement operators are optimal only for uniform weightings  taken from identified proper subsets, that is $\phi \in B(x,r) \subset [0, 2 \pi)$, where $B(x,r)$ denotes an open interval (open ball) centered at $x$ with radius $r \geq 0$. The numerical analysis identifies the following {\it{approximate}} intervals for $N \geq 4$, denoted by $B_{s}$, $B_{c,1}$ and $B_{c,2}$ (the complete graphs have two for $N \geq 4$) for the star and complete weighted graph states respectively,
\begin{multline}
\hspace*{-0.1cm}
    ~~~~~~~~~~~~~~~~~~~~~~~~~~B_s := B(\pi, r_s),\\ B_{c,1} := B(1, r_{c,1})\text{ and } B_{c,2} := B(2.5, r_{c,2}).
    \label{openb}
\end{multline}
The star and complete sub-classes, with weightings $\phi$ taken from intervals in Eq.~(\ref{openb}), are shown to yield superior precision results (see Fig.~\ref{B1}) when compared to other graph geometries with $|V|=N$. The corresponding radii values, $r_s$ and $r_c$, depend on $N$. A notable difference between the  sub-classes, is that the optimal weighting intervals for the star graph states share a common optimal center of $x_{s}= \pi$ (independent of system size $N$), while for complete graph states, the optimal weighting center $x_{c,1}$ and $x_{c,2}$ is a function of $N$, taking values in the intervals $[0.8,1.2]$ and $[2,3]$, respectively. For further analysis, based on the numerical results, we focus on $x_{c,1} = 1.04$, as it is practically relevant due to requiring a smaller phase $\phi$ and potentially easier to realize. Also, we find more optimal weighting intervals as $N$ increases, making it harder to choose specific ones. 

The points of discontinuity which are apparent in Fig.~\ref{B1}, occur at turning points of the expectation values of the identified optimal measurement operators, yielding singularities of the form $\frac{\cdot}{0}$ when evaluating Eq.~(\ref{D}). We consider the expectation values in more detail later in section IV, where we also present their analytic forms for arbitrary $N$.
\begin{figure}[t]
\hspace*{-0.4cm}
  \centering
  \begin{tabular}{ccc}
   \includegraphics[width=85mm,scale=0.995]{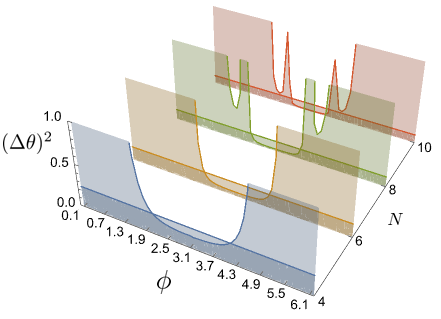}
  \end{tabular}
  \caption{Variance of the estimator for the weighted star graph state, using the measurement operator given in Eq.~(\ref{evenstar}), for $N \in \{4,6,8,10\}$ and $\phi_{ab} = \phi \in [0, 2 \pi)$ for all $\{a,b\} \in E$. The horizontal lines denote the SQL.}
  \label{A4s}
\end{figure}

\begin{figure}[t]
\hspace*{-0.15cm}
  \centering
  \begin{tabular}{ccc}
   \includegraphics[width=88mm,scale=0.995]{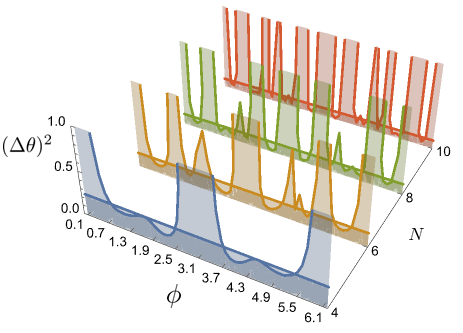}
  \end{tabular}
   \caption{Variance of the estimator for the weighted complete graph state, using the measurement operator given in Eq.~(\ref{allY}), for $N \in \{4,6,8,10\}$ and $\phi_{ab} = \phi \in [0, 2 \pi)$ for all $\{a,b\} \in E$. The horizontal lines denote the SQL.}
  \label{A4c}
\end{figure}

\begin{figure*}[t]
\hspace*{-0.1cm}
  \centering
  \begin{tabular}{ccc}
    \includegraphics[width=.94\textwidth]{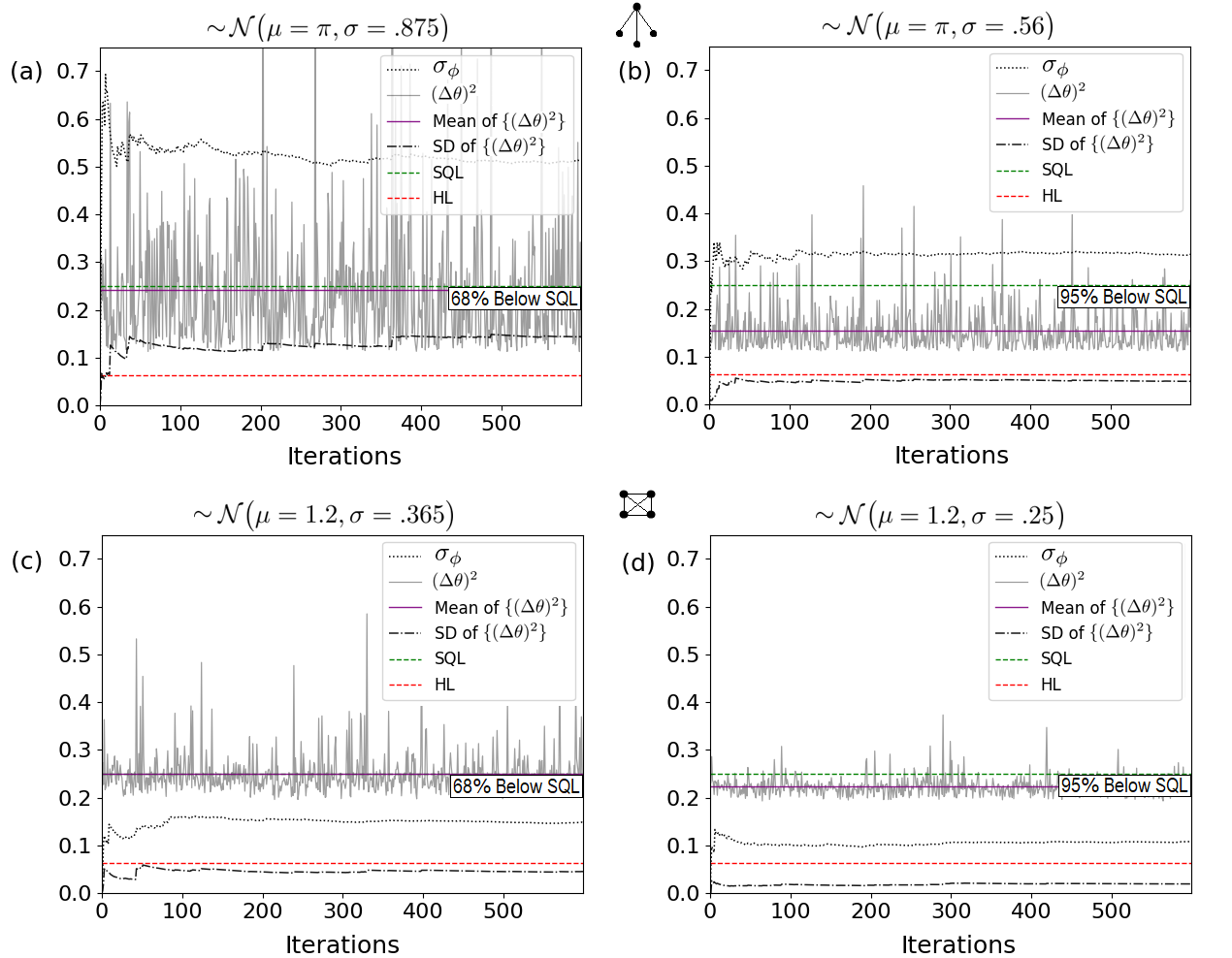}
  \end{tabular}
  \caption{Random Gaussian weighting distribution study (for $N=4$ and $\theta = 0.001$) over $600$ graph iterations, where $\phi_{ab} \sim \mathcal{N}(\mu, \sigma)$ for each $\{a,b\} \in E$. The Gaussian distribution standard deviation values $\sigma$ are chosen such that $68$\% and $95$\% (of the total $600$) of the optimal estimator variance values $(\Delta \theta)^2$ fall below the SQL ($1/N= 1/4)$. The SD of the set of weighting means per iteration is denoted by $\sigma_{\phi}$. (Top row) For the $4$-qubit star weighted graph state with $\mu = \pi,$ we find $\sigma = 0.875$ and $0.56$, respectively.  (Bottom row) For the $4$-qubit complete weighted graph state with $\mu = 1.2$, we find $\sigma = 0.365$ and $0.25$, respectively.}
  \label{randG}
\end{figure*}

%%%%%%%%%%%%%%%%%%%%%%%%%%%%
%%%%%%%%%%%%%%%%%%%%%%%%%%%%
%%%%%%%%%%%%%%%%%%%%%%%%%%%%
\subsection{Star Weighted Graph States}

 Consider the class of star weighted graph states, with optimal weighting interval center $x_{s}=\pi$, and where the optimal radius $r_{s}$ is a function of the system size $N$ (as shown in Fig.~\ref{A4s} for $N \in \{4, 6, 8, 10 \}$). For even $N \geq 4$, the optimal Pauli measurements which yield the results shown in Fig.~\ref{A4s}, are characterized as a function of $N$ by
\begin{align}
\hat{Y}\otimes \hat{Z}\otimes \hat{Y}^{\otimes N-3}\otimes\hat{Z}, \label{evenstar}   
\end{align}
where the first operator corresponds to the central qubit. Whereas for odd system size $N\geq3$, we find that the optimal Pauli measurement operator is 
\begin{align}
\hat{Y}\otimes \hat{Z}^{\otimes N-2}\otimes\hat{Y}.  
\label{oddstar}
\end{align}
The optimal measurement operators are found by a numerical search over all combinations of local Pauli operators of the form given by Eq.~(\ref{OpA}). The proofs are given in Appendix B.

%%%%%%%%%%%%%%%%%%%%%%%%%%%%
%%%%%%%%%%%%%%%%%%%%%%%%%%%%
%%%%%%%%%%%%%%%%%%%%%%%%%%%%
\subsection{Complete Weighted Graph States}

Now consider the class of complete weighted graph states, with optimal interval center $x_{c,1}\approx 1.$ (see Fig.~\ref{A4c}). The numerical search identifies the Pauli measurement that yields the optimal estimation results about $x_{c,1}$ as the all Pauli-$Y$ operator
\begin{align}
\hat{Y}^{\otimes N}:=\underbrace{\hat{Y}\otimes \cdot \cdot \cdot \otimes \hat{Y}}_{N}.   
\label{allY}
\end{align}

As demonstrated in Fig.~\ref{A4s} and Fig.~\ref{A4c}, the Pauli operators given by Eqs.~(\ref{evenstar}), (\ref{oddstar}) and (\ref{allY}) consistently produce estimator precision results below the classical SQL when choosing uniform weightings from the identified optimal intervals in Eq.~(\ref{openb}). The asymptotic behaviour as $N$ increases is related to the trend in the width of the region below the SQL in Fig.~\ref{A4s}. As $N$ increases the width changes less and less.   

%%%%%%%%%%%%%%%%%%%%%%%%%%%%
%%%%%%%%%%%%%%%%%%%%%%%%%%%%
%%%%%%%%%%%%%%%%%%%%%%%%%%%%
\subsection{Random Gaussian Weightings}

In Fig.~\ref{randG}, we demonstrate the robustness of the identified classes for $N=4$, in the sense that they consistently maintain favorable precision results even under imperfect preparation of the weighted graph state. We simulate this by modelling the set of edge weights $\{\phi_{ab}\}_{\{a,b\}\in E}$ as independent and identically distributed Gaussian random variables, $\phi_{ab}\sim\mathcal{N}(\mu,\sigma)$, where $\mu$ denotes the mean and $\sigma$ the standard deviation (SD) of the distribution. For a given weighted graph state, we then perform a fixed number of random samplings, or iterations, chosen to be $600$. This iteration number is chosen because both the SD of the set of weighting means
$
\left\{\sum_{\{a,b\}\in E}\phi_{ab}/M\right\}
$ (where $M = |E|$ is the number of edges and the corresponding SD is denoted by $\sigma_{\phi}$) and the SD of the corresponding set of optimal estimator variances $\big\{(\Delta \theta)^2\big\}$ reach stable values at this point (as shown in Fig.~\ref{randG}). As expected, $\sigma_{\phi} \to \sigma/\sqrt{M}$ for a sufficiently large iteration count.
 For the star and complete weighted graphs we find the Gaussian distribution $\sigma$ values which yield results where $68$\% (as shown in Fig.~\ref{randG}(a) and \ref{randG}(c)) and $95$\% (as shown in Fig.~\ref{randG}(b) and \ref{randG}(d)) of the total set of estimator values fall below the SQL. It is clear that the star graph is more robust to a distribution in the weights, although at a higher mean weighting of $\pi$.

\begin{figure}[t]
\hspace*{-0.25cm}
  \centering
  \begin{tabular}{ccc}
   \includegraphics[width=.49\textwidth]{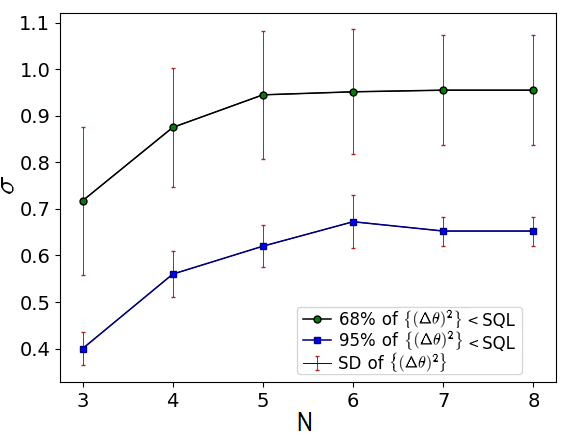}
  \end{tabular}
  \caption{The required Gaussian SD of IID weightings $\phi_{ab} \sim \mathcal{N}(\mu=\pi, \sigma)$, for star weighted graph states ($3 \leq N \leq 8$) over $600$ independent graph iterations, to yield $68$\% (and $95$\%) of optimal estimator variance values $\big\{(\Delta \theta)^2 \big\}$ below the SQL. The error bars denote the SD of the optimal set $\big\{(\Delta \theta)^2 \big\}$.}
  \label{A5}
\end{figure}

As previously mentioned, the class of weighted star graphs has the property that the optimal weighting center $x = \pi$ is independent of $N$, this warrants an extended robustness study to higher $N$ with $\mu = \pi$, the results of which are shown in Fig.~\ref{A5}. 

%%%%%%%%%%%%%%%%%%%%%%%%%%%%
%%%%%%%%%%%%%%%%%%%%%%%%%%%%
%%%%%%%%%%%%%%%%%%%%%%%%%%%%
%%%%%%%%%%%%%%%%%%%%%%%%%%%%
\section{Analytic Forms}

The above study provides interesting insights into the quantum advantage that small weighted graph states can provide for quantum sensing. It thus motivates the study of larger $N$. For this, we study simple cases that admit an analytical formula for larger $N$.

For the star and complete weighted graph states, we provide analytic forms of the QFI and expectation values of the measurements of $\hat{A}$, from which the corresponding form of the estimator variance follows easily using Eq.~(\ref{D}). The analytic forms presented assume uniform weightings, that is $\phi_{ab}=\phi \in [0, 2\pi)$ for all $\{a,b\}\in E$. 

\begin{figure}[tbph]
\hspace*{0cm}
  \centering
  \begin{tabular}{ccc}
\includegraphics[width=85mm,scale=0.99]{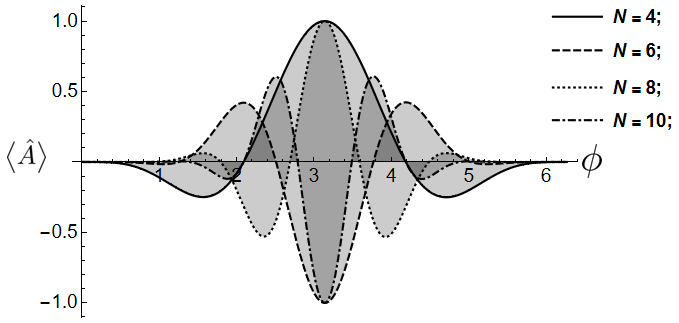}
  \end{tabular}
  \caption{The expectation value $\langle \hat{A} \rangle$ of the optimal Pauli measurement operator given in Eq.~(\ref{evenstar}) for the star weighted graph state with $N \in \{4,6,8,10\}$ and $\theta = 0.001$.
}
  \label{exps2}
\end{figure}

%%%%%%%%%%%%%%%%%%%%%%%%%%%%
%%%%%%%%%%%%%%%%%%%%%%%%%%%%
%%%%%%%%%%%%%%%%%%%%%%%%%%%%
\subsection{Analytic Forms of Star Weighted Graph States}

%%%%%%%%%%%%%%%%%%%%%%%%%%%%
%%%%%%%%%%%%%%%%%%%%%%%%%%%%
\subsubsection{Quantum Fisher Information}
The QFI of an $N$-qubit weighted star graph state, with uniform weighting $\phi$, is given by

\begin{multline}
\hspace*{0cm}
\mathcal{Q}(\phi, N)=\frac{1}{16}(N-2)(N-1)e^{-2i\phi}(e^{i \phi}-1)^4 
\\ -\frac{(N-1)e^{-i \phi}(e^{i \phi}-1)^2\big(e^{i \phi}(1+e^{-i \phi})^N+(1+e^{i \phi})^N\big)}{(1+e^{i \phi})2^{N+1}}
\\-\frac{1}{4}\bigg(\bigg(\frac{1}{2}+\frac{e^{-i \phi}}{2}\bigg)^{N-1}+\bigg(\frac{1}{2}+\frac{e^{i \phi}}{2}\bigg)^{N-1}\bigg)^2
\\+ (N-1)\bigg(1-\frac{1}{16}e^{-2i \phi}(1+e^{i \phi})^4\bigg)+1.
\label{QFIs}
\end{multline}
For fully-weighted star graph states, that is $\phi = \pi$, the QFI simplifies to that given in Ref. \cite{shettell2020graph},
\begin{align}
\mathcal{Q}(\pi, N) = (N-1)^2+1.
\end{align}
The proof is given in Appendix B.

%%%%%%%%%%%%%%%%%%%%%%%%%%%%
%%%%%%%%%%%%%%%%%%%%%%%%%%%%
\subsubsection{Expectation Value}
We infer from the numerical results that operators of the form Eqs.~(\ref{evenstar}) and (\ref{oddstar}), yield favorable metrological results. For an even $N$-qubit weighted star graph state, the analytic form of the expectation value, for the measurement operator given by Eq.~(\ref{evenstar}), reads as 
% \begin{multline}
\begin{equation}
\begin{split}
   \big\langle \hat{A} \big\rangle = -\frac{2^{-N}\cos \theta (e^{-i \theta})\big[(-1+e^{i \phi})(-ie^{i \theta})\big]^N}{e^{i \phi}-1} \\-
    2^{-N}i\cos \theta \big[e^{-i \phi}(-1+e^{i \phi})(-i e^{-i \theta})\big]^{N-1}\\+
    2^{-N}\sin \theta \tan^{2}\theta\big[-i e^{-i \phi}(e^{2 i \phi}-1)\cos \theta\big]^{N-1}.
    \label{exps}
% \end{multline}
\end{split}
\end{equation}
The proof of this and the corresponding expression for odd $N$, can be found in Appendix B. 

Fig.~{\ref{exps2}} shows the expectation value given by Eq.~(\ref{exps}) for $N\in \{4,6,8,10\}$ with $\theta = 0.001$. Due to the Pauli matrices being involutory, it follows that $\langle \hat{A}^2\rangle = 1$ when restricting to Pauli measurement operators of the form given by Eq.~$(\ref{OpA})$. Thus the analytic form of the estimator variance $(\Delta \theta )^{2}$ follows easily, since from Eq.~(\ref{exps}) we obtain the required expressions of $(\Delta A)^2 = 1 -~\langle \hat{A} \rangle^2$ and $\big|\partial_{\theta} \big\langle \hat{A} \big\rangle\big|^2$.

Fig.~\ref{sps} shows the estimator variance $(\Delta \theta)^2$ for weighted star graph states, given the measurement operator in Eq.~(\ref{evenstar}),  for even $N$ and $\theta =~0.001$. It is apparent that although differing in interval size, there is a common minimum estimator region (or trough) centered about the weighting $\phi = \pi$ which occurs for $4\leq N \leq 100$. This is consistent with our previous numerical results. The number of troughs on either side of $\pi$ is observed to increase symmetrically with increasing system size $N$.

%%%%%%%%%%%%%%%%%%%%%%%%%%%%
%%%%%%%%%%%%%%%%%%%%%%%%%%%%
%%%%%%%%%%%%%%%%%%%%%%%%%%%%
\subsection{Analytic Forms of Complete Weighted Graph States}

%%%%%%%%%%%%%%%%%%%%%%%%%%%%
%%%%%%%%%%%%%%%%%%%%%%%%%%%%
\subsubsection{Quantum Fisher Information}
The QFI of an $N$-qubit complete weighted graph state, with uniform weighting $\phi$, is given by
\begin{multline}
\mathcal{Q}(\phi,N)=
N\bigg[1+2^{-N}(N-1)\\\times \bigg(2^{N-1}+\frac{e^{i \phi}\big(e^{i 2 \phi}(1+e^{-2i\phi})^{N}+(1+e^{2i\phi})^{N}\big)}{(1+e^{2i \phi })^2}\bigg) \\- \frac{4^{-N}N\big(e^{i \phi}(1+e^{-i \phi})^N+(1+e^{i \phi})^N\big)^2}{(1+e^{i \phi})^2} \bigg].
\label{QFIc}
\end{multline}
The proof is given in Appendix B.
For fully-weighted complete graph states with $\phi=\pi$, Eq.~(\ref{QFIc}) simplifies to that given in Ref.~\cite{shettell2020graph},
\begin{align}
\mathcal{Q}(\pi, N) = N.
\end{align}

\begin{figure}[t]
\hspace*{-0.2cm}
  \centering
  \begin{tabular}{ccc}
  \includegraphics[width=89mm,scale=0.99]{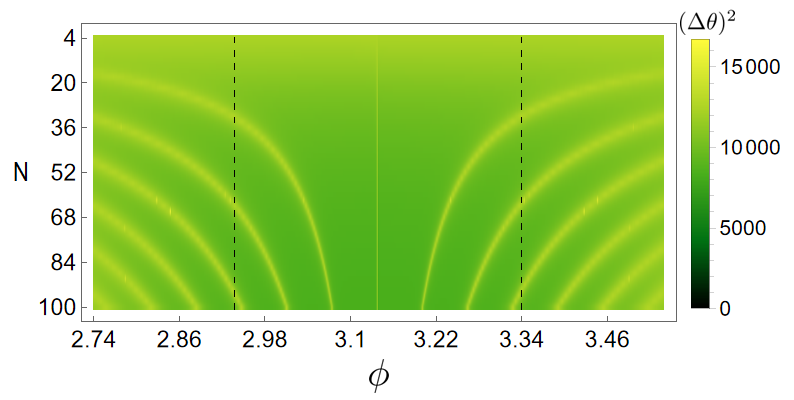}
  \end{tabular}
  \caption{The star weighted graph state robustness study for $4\leq N \leq 100$ (for even $N$), where $\phi \in B_{s} := B(\pi, 0.4)$. The dashed lines demarcate the end points of the weighting interval $B(\pi, 0.2)$.}
  \label{sps}
\end{figure}

\begin{figure}[t]
\hspace*{-0.2cm}
  \centering
  \begin{tabular}{ccc}
   \includegraphics[width=89mm,scale=0.99]{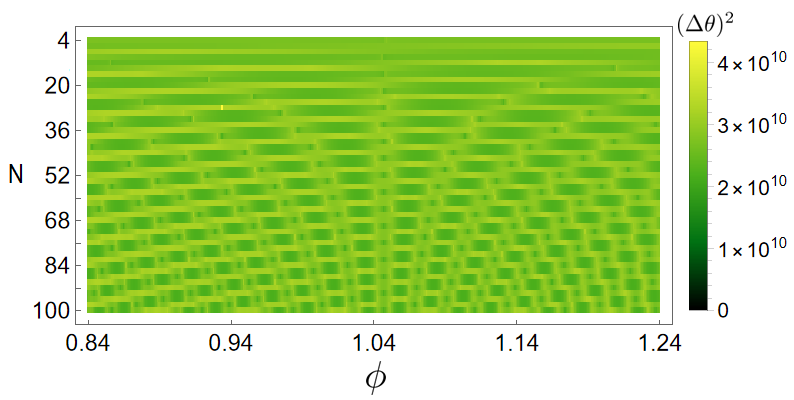}
  \end{tabular}
  \caption{The complete weighted graph state robustness study for $4\leq N \leq 100$ (for even $N$), where $\phi \in B_{c,1} := B(1, r_{c,1}=~0.2)$.}
  \label{spc}
\end{figure}

%%%%%%%%%%%%%%%%%%%%%%%%%%%%
%%%%%%%%%%%%%%%%%%%%%%%%%%%%
\subsubsection{Expectation Value}
From the numerical analysis results for complete weighted graphs, we infer that the all Pauli-$Y$ measurement operator in Eq.~(\ref{allY}) yields the optimal precision results. 
The analytic form of the corresponding expectation value, for both even and odd $N$, reads as 
\begin{multline}
    \big \langle \hat{A} \big\rangle = \sum_{m=1}^{N}i^{m}2^{-N}(-1)^{N-m}\binom{N}{m}\cos^m\theta\sin^{N-m}\theta \\\times \sum_{a=0}^{N-m}\sum_{b=0}^{m}(-1)^{a+m-b}\binom{m}{b}e^{i \phi \big(a+\frac{m-1}{2}\big)(m-2b)}\binom{N-m}{a}.
    \label{ExpComp}
\end{multline}

The proof is given in Appendix B.
Again, we have $\langle \hat{A}^2 \rangle = 1$. The analytic form of the estimator variance $(\Delta \theta )^{2}$ follows easily, since from Eq.~(\ref{ExpComp}) we obtain the required analytic expressions of $(\Delta A)^2$ and $\big|\partial_{\theta} \big\langle \hat{A} \big\rangle\big|^2$.

Fig.~\ref{spc} shows the estimator variance $(\Delta \theta)^2$ for the complete weighted graph states for even $N \in \{4,...,100\}$, given the measurement operator in Eq.~(\ref{allY}), with $\phi~\in B(1.04, r_{c,1}=0.2)$ and $\theta=0.001$. As opposed to the results of the star weighted graphs presented in Fig.~\ref{sps}, for the complete weighted graphs we observe more troughs with shorter interval lengths. The complete weighted graphs exhibit more troughs in $(\Delta \theta)^2$, indicating more distinct $\phi$-regions that locally optimize the estimation precision. These troughs occur over shorter $\phi$-intervals, suggesting reduced robustness to edge-weight miscalibration compared with the star subclass, whose favorable regions are broader. 

%%%%%%%%%%%%%%%%%%%%%%%%%%%%
%%%%%%%%%%%%%%%%%%%%%%%%%%%%
%%%%%%%%%%%%%%%%%%%%%%%%%%%%
%%%%%%%%%%%%%%%%%%%%%%%%%%%%
\section{Analysis for $N > 100$}

\begin{figure*}[tbph]
\hspace*{-0.15cm}
  \centering
  \begin{tabular}{ccc}
    \includegraphics[width=1.02\textwidth]{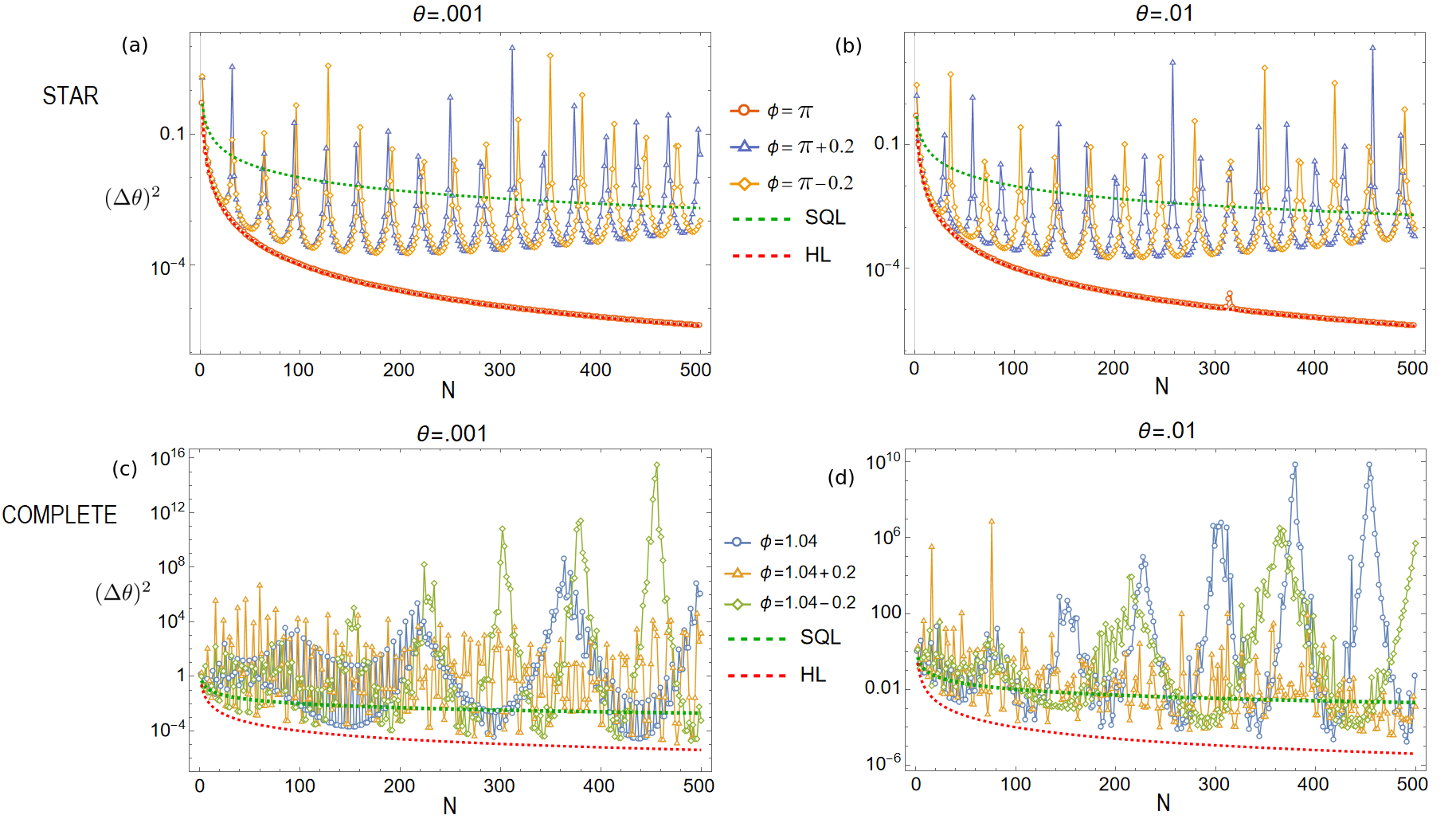}
  \end{tabular}
  \caption{The estimator variance $(\Delta \theta)^2$ for even $N \in \{2,...,500\}$ and $\theta \in \{0.001, 0.01\}$, of the star (top) and complete (bottom) weighted graph states centered about optimal uniform weightings $\phi = \pi$ and $1.04$, respectively (with logarithmic scaling).}
  \label{lp}
\end{figure*}

The analytic forms allow us to further investigate the efficacy of the identified sub-classes for larger system size $N$.
Fig.~\ref{lp} shows the estimator variance $(\Delta \theta)^2$ for the star and complete weighted graph states for even $N \in \{2,...,500\}$ and $\theta \in \{0.001, 0.01\}$. The plots for $\theta = 0.001$ for the star and complete weighted graph, i.e., Fig.~\ref{lp}(a) and (c) are essentially vertical cross sections at a specified $\phi$, of Fig.~\ref{sps} and \ref{spc}, respectively, but for up to $N=500$.  In Fig.~\ref{lp}(a) and (b) we consider the star weighted graph state results. For $N \gtrapprox 4$ and $\phi = \pi$, the optimal estimator variance lies close to the HL. Whereas for imperfect weightings $\phi = \pi \pm 0.2$, the majority of the set of optimal estimator variance values $\{(\Delta \theta)^2\}$ are dispersed between the SQL and HL bounds. For $\theta = 0.001$ and $\theta = 0.01$, some values of $N$ yield optimal estimator variance values above the SQL, but most lie well below it.
 As $N$ increases the estimator variance values corresponding to the imperfect weightings tend to the SQL. 

In Fig.~\ref{lp}(c) and (d) we consider the class of complete weighted graph states. There is a less discernible difference when comparing the numerical results for the optimal uniform weighting $\phi = 1.04$, with that of the imperfect weightings $\phi = 1.04 \pm 0.2$. A noted feature when compared to the star weighted graph results, is that even the imperfect uniform weightings $\phi = 1.04 \pm 0.2$, although intermittently, continue yielding values close to the HL bound for increasing system size $N$. 

The similarity between the star graph results for the non-optimal uniform weightings $\phi = \pi \pm 0.2$, suggests a symmetry about the optimal weighting $\pi$. This symmetry is not observed for the complete graphs about optimal weighting $\phi = 1.04$, where there is a more discernible difference between the results for non-optimal weightings $\phi = 1.04 \pm 0.2$. The results presented in Fig.~\ref{lp} appear relatively stable to an order of magnitude increase of the encoded unitary phase $\theta \in \{0.001, 0.01\}$ when considering the general patterns and trends, but there is a distinct difference observed for both classes of weighted graph states.

%%%%%%%%%%%%%%%%%%%%%%%%%%%%
%%%%%%%%%%%%%%%%%%%%%%%%%%%%
%%%%%%%%%%%%%%%%%%%%%%%%%%%%
%%%%%%%%%%%%%%%%%%%%%%%%%%%%
\section{Discussion}

In this work we considered weighted graph states as a resource
for the canonical phase estimation task in metrology. We successfully identified metrologically relevant sub-classes which produce a favorable precision below the SQL; approaching the HL. More specifically, by considering classifications by their unique graph geometry, the star and complete weighted graph states were identified as consistently producing superior precision results when compared to other graph geometries.

Furthermore, we presented analytic forms of the QFI and estimator variance of the optimal local Pauli measurements for arbitrary $N$. Using these results, we demonstrated that even with the restricted class of local measurement operators, favorable metrological results are possible and maintained for large system sizes. For complete weighted graphs we also found that the weights can be much reduced from $\pi$ and still give a quantum advantage. This makes them more accessible from a practical point of view. 

Our robustness study then showed that imperfect weightings, within a certain range about the optimal weightings, are still sufficient to produce favorable results below the SQL. 

An important consideration when characterizing metrologically relevant quantum states, is their ability to tolerate noise, in the sense that a distinct quantum advantage is maintained when including quantum decoherence. Further analysis is needed to study the efficacy of these sub-classes when including decoherence. The robustness in state preparation hints at a more general robustness to noise from the environment. But to adequately address this topic, a dedicated numerical and theoretical analysis is needed.  

%%%%%%%%%%%%%%%%%%%%%%%%%%%%
%%%%%%%%%%%%%%%%%%%%%%%%%%%%
%%%%%%%%%%%%%%%%%%%%%%%%%%%%
%%%%%%%%%%%%%%%%%%%%%%%%%%%%
\section{Acknowledgements}
We acknowledge the National Research Foundation, the South African Council for Scientific and Industrial Research, and the Department of Science and Innovation through the South African Quantum Technology Initiative for funding this project. S.K.O. acknowledges support from Air Force Office of Scientific Research (AFOSR) Multidisciplinary University Research Initiative (MURI) Award No. FA9550-21-1-0202.

%%%%%%%%%%%%%%%%%%%%%%%%%%%%
%%%%%%%%%%%%%%%%%%%%%%%%%%%%
%%%%%%%%%%%%%%%%%%%%%%%%%%%%
%%%%%%%%%%%%%%%%%%%%%%%%%%%%
\appendix
% \sectionfont{\MakeUppercase}

%%%%%%%%%%%%%%%%%%%%%%%%%%%%
%%%%%%%%%%%%%%%%%%%%%%%%%%%%
%%%%%%%%%%%%%%%%%%%%%%%%%%%%
\section{Point Estimation}

 The point estimation problem attempts to find a function $\hat{\theta}(X_1,...,X_n)$, where $X_1,...,X_n$ is a random sample chosen from a cumulative distribution function (CDF), denoted by $F_{\theta}$, that accurately approximates an unknown distribution parameter $\theta$. An example estimator is the sample mean, $\hat{\theta}:=\frac{1}{n}\sum_{i}X_{i}$. The CDF admits a probability mass function, for the discrete case, and a probability density function, for the continuous case. A point estimate of $\theta$, is the value of the function $\hat{\theta}:=\hat{\theta}(X_1,...,X_n)$ for a single sample, with values restricted to a given parameter space $\Theta \subset \mathbb{R}^d,~ d\geq1$. In a quantum–mechanical setting, the random variables $X_i$ can be regarded as outcomes of repeated measurements of a Hermitian observable $\hat{A}$ on identically prepared probe states. The estimator $\hat{\theta}(X_1,\dots,X_n)$ is then a classical post-processing of these measurement outcomes, so its variance is directly tied to the measurement statistics of $\hat{A}$. 
 
 Estimators are not unique, and as such, we seek to identify the optimal estimator or set of estimators for the defined estimation task. To that end, it is necessary to set a standard for comparing estimators. Consider the well-known mean-square error (MSE) \cite{pishro2016introduction},  
\begin{equation}
\text{MSE}_{\theta}(\hat{\theta}):=E_{\theta}\big\{(\hat{\theta}-\theta)^2\big\},
    \label{MSE}
\end{equation}
where (\ref{MSE}) denotes the mean of the squared difference between the estimator value $\hat{\theta}(X_1,...,X_n)$ and the true value $\theta$, sampled from $F_{\theta}$. For an unbiased estimator, the MSE is equal to the estimator variance. 

Note that for our purposes, we do not explicitly evaluate the estimator $\hat{\theta}(X_1,...,X_n)$, as required by (\ref{MSE}), but instead consider the estimator variance as a measure of the estimation precision (see Eq.~(\ref{D})). It follows that the precision in estimation depends on the variance of the chosen measurement operator $\hat{A}$ (which we seek to minimize), and the sensitivity of the expectation value $\langle \hat{A} \rangle$ to changes in $\theta$ (which we seek to maximize).
%%%%%%%%%%%%%%%%%%%%%%%%%%%%
%%%%%%%%%%%%%%%%%%%%%%%%%%%%
%%%%%%%%%%%%%%%%%%%%%%%%%%%%
\section{Proofs of IV. Analytic Forms }

%%%%%%%%%%%%%%%%%%%%%%%%%%%%
%%%%%%%%%%%%%%%%%%%%%%%%%%%%
\subsection*{Star weighted graph states}

%%%%%%%%%%%%%%%%%%%%%%%%%%%%
\subsubsection{QFI}

\begin{proof}
The state vector of an $N$-qubit star weighted graph state reads as  
\begin{align}
    |\psi_{N}\rangle=\frac{1}{\sqrt{2}}|0\rangle\otimes|+\rangle^{\otimes N-1}+\frac{1}{\sqrt{2}}|1\rangle\otimes|+_{\phi}\rangle^{\otimes N-1},
    \label{star}
\end{align}
where $|+\rangle := \frac{1}{\sqrt{2}}(|0\rangle + |1\rangle)$ and $|+_{\phi}\rangle:=\frac{1}{\sqrt{2}}(|0\rangle + e^{i \phi}|1\rangle)$. 
Let $\rho = | \psi_N\rangle \langle \psi_N|$. From Eq.~(\ref{star}) it follows that
% \begin{equation}
\begin{multline}
\text{Tr}(\hat{X}_{j}\rho) \\= \begin{cases}
     \frac{1}{2}\bigg(\frac{1}{2}+\frac{1}{2}e^{i \phi}\bigg)^{N-1}+\frac{1}{2}\bigg(\frac{1}{2}+\frac{1}{2}e^{-i \phi}\bigg)^{N-1}, & \text{if } j = 1,\\
      \frac{1}{2}+\frac{1}{2}\big(\frac{1}{2}e^{i \phi}+\frac{1}{2}e^{-i \phi} \big), & \text{if } j \neq 1.
         \end{cases}
         \label{s1}
% \end{equation}
\end{multline}
For $k=1,~j\neq1$ (or equivalently $j=1,~k\neq1$) we find that     $\text{Tr}(\hat{X}_{j}\hat{X}_{1} \rho )=   \text{Tr}(\hat{X}_{1} \rho)$. For $j=k=1$ it follows that $\text{Tr}(\hat{X}_{j}\hat{X}_{k}\rho) = \text{Tr}(\rho) = 1$. 
% \begin{equation}
%      \text{Tr}(\hat{X}_{j}\hat{X}_{1} \rho )=   \text{Tr}(\hat{X}_{j} \rho).
%           \label{s2}
% \end{equation}
Finally, for $j \neq 1,~k\neq 1$ we find that 
\begin{equation}
\text{Tr}( \hat{X}_{j}\hat{X}_{k} \rho) = \begin{cases}
            1, & \text{if } j = k,\\
      \frac{1}{2} + \frac{1}{2}\big(\frac{1}{2}e^{i \phi}+\frac{1}{2}e^{-i \phi} \big)^2, & \text{if } j \neq k.
         \end{cases}
               \label{s3}
\end{equation}
By utilizing Eqs.~(\ref{s1}) and (\ref{s3}) to evaluate the summation given in Eq.~(\ref{QFI2}), it follows that the QFI of an $N$-qubit star weighted graph state, after simplification, is given by Eq.~(\ref{QFIs}).
\end{proof}

%%%%%%%%%%%%%%%%%%%%%%%%%%%%
\subsubsection{Expectation value}
\begin{proof}From the numerical analysis (see Fig.~\ref{B1} and Fig.~\ref{A4s}), we find that the optimal local Pauli measurements are given by Eqs.~(\ref{evenstar}) and (\ref{oddstar}). Ultimately, we seek to derive an analytic form of the associated estimator variance given by Eq.~(\ref{D}), to that end, it is sufficient to derive an analytic form of the expectation value of the measurement operator. For odd system size $N$, it is instructive to first consider the $3$-qubit star weighted graph state, that is, $$|\psi_{3}\rangle=\frac{1}{\sqrt{2}}|0\rangle \otimes |+\rangle^{\otimes 2}+\frac{1}{\sqrt{2}}|1\rangle \otimes |+_{\phi}\rangle^{\otimes 2},$$ 
where the optimal local Pauli measurement, for $N=3$, is of the form Eq. (\ref{oddstar}), that is, 
\begin{align}
\hat{A}:=\hat{Y}\otimes\hat{Z}\otimes\hat{Y}.
\end{align}
In the Heisenberg picture, assuming the Hamiltonian given by Eq.~(\ref{C}), the measurement operator evolution reads as
\begin{gather} 
\hspace*{-0.75cm}
\hat{A}(\theta) = e^{\big(i\frac{\theta}{2}\sum_{j=1}^{N=3}\hat{X}_{j}\big)}\hat{A}e^{\big(-i\frac{\theta}{2}\sum_{j=1}^{N=3}\hat{X}_{j}\big)} \nonumber =\\ 
(\cos \theta \hat{Y}-\sin \theta \hat{Z}) \otimes (\cos \theta \hat{Z}+\sin \theta \hat{Y}) \otimes (\cos \theta \hat{Y}-\sin \theta \hat{Z}),
\end{gather}
where the last equation utilizes the matrix generalization of Euler's formula. 
Now consider the expectation value with respect to the $3$-qubit star weighted graph state
\begin{equation}
\hspace*{-0.22cm}
\begin{split}
\langle &\psi_{3} | \hat{A}(\theta)|\psi_3\rangle  \\&= \frac{i}{2}\cos \theta \bigg[\frac{1}{2}(\cos \theta - i \sin \theta)+\frac{e^{-i \phi}}{2}(i \sin \theta-\cos \theta)\bigg]
\\& \times \bigg[\frac{e^{-i \phi}}{2}(i\cos \theta + \sin \theta)-\frac{1}{2}(i \cos \theta+\sin \theta)\bigg] \\ & -\frac{i}{2}\cos \theta \bigg[\frac{1}{2}(\cos \theta - ie^{i \phi} \sin \theta)+\frac{1}{2}(i \sin \theta-e^{i \phi}\cos \theta)\bigg] 
\\& \times \bigg[\frac{1}{2}(i\cos \theta + e^{i \phi} \sin \theta)-\frac{1}{2}(i e^{i \phi} \cos \theta+\sin \theta)\bigg]
\\& +\frac{1}{2}\sin \theta \bigg[\frac{1}{2}(\cos \theta - i e^{i \phi} \sin \theta)+\frac{e^{-i \phi}}{2}(i \sin \theta-e^{i \phi}\cos \theta)\bigg]  
\\& \times \bigg[\frac{e^{-i \phi}}{2}(i\cos \theta + e^{i \phi} \sin \theta)-\frac{1}{2}(i e^{i \phi} \cos \theta+\sin \theta)\bigg].
\label{Expstar}
\end{split}
\end{equation}

Exploiting the graph symmetry of the star weighted graph states, we can easily generalize the above for the odd $N$-qubit expectation value, by raising the exponent of the first square bracket $\big[ \cdot \big]$ of each of the three terms in Eq.~(\ref{Expstar}), from $1$ to $N-2$. Similarly, for an even $N$-qubit star weighted graph state, given the optimal Pauli measurement operator in Eq.~(\ref{evenstar}), the expectation value reads as 
\begin{equation}
\hspace*{-0.18cm}
\begin{split}
&\langle \psi_{N} | \hat{A}(\theta)|\psi_{N}\rangle  \\&= \frac{i}{2}\cos \theta \bigg[\frac{1}{2}(\cos \theta - i \sin \theta)+\frac{e^{-i \phi}}{2}(i \sin \theta-\cos \theta)\bigg]^{2} 
\\& \times \bigg[\frac{e^{-i \phi}}{2}(i\cos \theta + \sin \theta)-\frac{1}{2}(i \cos \theta+\sin \theta)\bigg]^{N-3} \\ & -\frac{i}{2}\cos \theta \bigg[\frac{1}{2}(\cos \theta - ie^{i \phi} \sin \theta)+\frac{1}{2}(i \sin \theta-e^{i \phi}\cos \theta)\bigg]^{2} 
\\& \times \bigg[\frac{1}{2}(i\cos \theta + e^{i \phi} \sin \theta)-\frac{1}{2}(i e^{i \phi} \cos \theta+\sin \theta)\bigg]^{N-3} 
\\& +\frac{1}{2}\sin \theta \bigg[\frac{1}{2}(\cos \theta - i e^{i \phi} \sin \theta)+\frac{e^{-i \phi}}{2}(i \sin \theta-e^{i \phi}\cos \theta)\bigg]^{2}  
\\& \times \bigg[\frac{e^{-i \phi}}{2}(i\cos \theta + e^{i \phi} \sin \theta)-\frac{1}{2}(i e^{i \phi} \cos \theta+\sin \theta)\bigg]^{N-3},
\label{Expstar2}
\end{split}
\end{equation}
where Eq.~(\ref{Expstar2}) simplifies yielding Eq.~(\ref{exps}).
\end{proof}
\subsection*{Complete weighted graph states}
\subsubsection{QFI}
\begin{proof}
We seek to derive the QFI, given by Eq.~(\ref{QFIc}), of an $N$-qubit complete weighted graph state.
First consider the following useful equivalent form of a complete weighted graph state (see Eq.~(\ref{E})),
\begin{equation}
\begin{split}
    |\psi\rangle &= \prod_{\substack{l,j=1\\ l< j}}^{N}CZ^{\phi_{lj}}|+\rangle^{\otimes N}
    \\& = 2^{-N/2} \prod_{\substack{l< j}}^{N}CZ^{\phi_{lj}}\sum_{\underbar{k}}|k_1\cdot\cdot\cdot k_N\rangle 
    \\& = 2^{-N/2}\sum_{\underbar{k}} \prod_{\substack{l< j}}^{N}e^{i\phi_{lj}(k_{l}\wedge k_{j})}|k_1\cdot\cdot\cdot k_N\rangle, 
    \label{WGS2}
\end{split}
\end{equation}
where $CZ^{\phi_{lj}}$ denotes a controlled-$\phi$ phase gate $\hat{U}(\phi)=\text{diag}(1,1,1,e^{i \phi})$ acting on the $l$-th and $j$-th qubits, and $\underbar{k}$ denotes the set of binary permutations of $|k_1\cdot\cdot\cdot k_N\rangle$, with $k_i \in \{0,1\}$. The expression for the QFI in Eq.~(\ref{QFI2}) involves expectation values of $\hat{X}_{m}$ and $\hat{X}_{m}\hat{X}_{w}$. We start with the $\hat{X}_{m}$ terms. Consider the following transformation 
\begin{equation}
\begin{split}
|\psi_{m}\rangle&:=\hat{X}_{m}|\psi\rangle \\& =  2^{-N/2}\sum_{\underbar{k}} \prod_{\substack{l< j}}^{N}e^{i\phi_{lj}(k_{l}\wedge k_{j})}\hat{X}_{m}|k_1\cdot\cdot\cdot k_N\rangle
\\& =  2^{-N/2}\sum_{\underbar{k}} \prod_{\substack{l< j}}^{N}e^{i\phi_{lj}\big((k_{l}\oplus \delta_{ml})\wedge (k_{j}\oplus \delta_{mj})\big)}|k_1\cdot\cdot\cdot k_N\rangle,
\end{split}
\end{equation}
where $\delta$ denotes the Kronecker delta, $\wedge$ denotes the logical conjunction operation and $\oplus$ denotes modulo-2 addition. Therefore,
\begin{equation}
\begin{split}
\langle \psi|\psi_{m}\rangle&=2^{-N}\sum_{\underbar{k}} \prod_{\substack{l< j}}^{N}e^{i\phi_{lj}\big((k_{l}\oplus \delta_{ml})\wedge (k_{j} \oplus \delta_{mj})-(k_{l}\wedge k_{j})\big)}.
\label{psim}
\end{split}
\end{equation}
 Similarly, we define $|\psi_{mw}\rangle := \hat{X}_{m}\hat{X}_{w}|\psi\rangle$, it then follows that 
\begin{multline}
\langle \psi|\psi_{mw}\rangle \\~~~=2^{-N}\sum_{\underbar{k}} \prod_{\substack{l< j}}^{N}e^{i\phi_{lj}\big((k_{l}\oplus \delta_{ml}\oplus \delta_{wl})\wedge (k_{j} \oplus \delta_{mj} \oplus \delta_{wj})-(k_{l}\wedge k_{j})\big)}.
\label{psiw}
\end{multline}
Note that $m=w$ implies  $\langle \psi|\psi_{mm} \rangle = \langle\psi|\psi\rangle=1$, which occurs $N$ times. Whereas for $m\neq w$, there are $2\binom{N}{2}$ unique pairwise combinations. For an $N$-qubit complete weighted graph state, we consider the elements of the corresponding adjacency matrices are $\phi_{lj} = \phi$ for all $\{l,j\} \in E$. Hence, Eq.~(\ref{psim}) is identical for all $m \in \{1,...,N \}$ and Eq.~(\ref{psiw}) is identical for all $m,w \in \{1,...,N \}$.   

By evaluating Eq.~(\ref{psim}) for increasing system size $N$, we find that the coefficients of terms in Eq.~(\ref{psim}) correspond to the rows of Pascal's Triangle, more specifically, $\binom{N-1}{k}$. Thus, for the class of complete weighted graph states, Eq.~(\ref{psim}) takes the simplified form 
\begin{equation}
\begin{split}
\langle \psi|\psi_{m}\rangle&=2^{-N}\sum_{k=0}^{N-1}\binom{N-1}{k}\big(e^{i \phi k}+e^{- i \phi k}\big).
\label{psim2}
\end{split}    
\end{equation}
Similarly, for Eq.~(\ref{psiw}) we find the simplified form of
\begin{multline}
\langle \psi|\psi_{mw}\rangle
\\~~~~=2^{-N}\bigg[2^{N-1}+\sum_{k=0}^{N-2}\binom{N-2}{k}\big(e^{i \phi (2k+1)}+e^{- i \phi (2k+1)}\big)\bigg].
\label{psiw2}
\end{multline}
By substituting Eqs.~(\ref{psim2}) and (\ref{psiw2}) into Eq.~(\ref{QFI2}) and simplifying, we obtain the analytic form given by Eq.~(\ref{QFIc}). 
\end{proof}

%%%%%%%%%%%%%%%%%%%%%%%%%%%%
\subsubsection{Expectation value} 
\begin{proof}  First consider the following useful identity. Since $k_{l} \wedge k_{j} = k_{l}k_{j}$ (scalar multiplication), it follows that
\begin{equation}
\begin{split}
    \sum_{\underbar{k}} \prod_{\substack{l< j}}^{N}e^{i\phi(k_{l}\wedge k_{j})} &=\sum_{\underbar{k}} \prod_{\substack{l< j}}^{N}e^{i\phi k_{l} k_{j}}
    \\& =\sum_{k=0}^{N}\sum_{\substack{\underbar{k} \text{ with }\\ k \text{ ones}}} \prod_{\substack{l< j}}^{N}e^{i\phi k_{l} k_{j}}
    \\& = \sum_{k=0}^{N}\binom{N}{k}e^{i \phi \binom{k}{2}}. 
    \label{Cp1}
\end{split}
\end{equation}
Now consider an $N$-qubit complete weighted graph state $|\psi\rangle$. By definition all qubits are connected to each other, with uniform interaction weighting $\phi$, thus $\phi_{lj} = \phi$ for all $\{l,j\}\in E$. Since the Pauli-$Y$ operator maps $|0\rangle \mapsto~i|1\rangle$ and $|1\rangle \mapsto -i|0\rangle$, we have the following all Pauli-$Y$ transformation, $|\psi_{Y}\rangle := \hat{Y}^{\otimes N}|\psi\rangle$, with
\begin{equation}
\hspace*{-0.5cm}
\begin{split}
&~~|\psi_{Y}\rangle =  2^{-N/2}\sum_{\underbar{k}} \prod_{\substack{l< j}}^{N}e^{i\phi(k_{l}\wedge k_{j})}\underbrace{\hat{Y}\otimes \cdot \cdot \cdot \otimes \hat{Y}}_{N}|k_1\cdot\cdot\cdot k_N\rangle
\hspace*{-0.5cm}
\\& =  2^{-N/2}(i)^{N}\sum_{\underbar{k}}(-1)^{\sum_{s}^{N}k_s} \prod_{\substack{l< j}}^{N}e^{i\phi(k_{l}\wedge k_{j})}|k_1\oplus 1\cdot\cdot\cdot k_N \oplus 1\rangle. 
\end{split}
\end{equation}
The following is an equivalent form of the complete weighted graph state 
\begin{equation}
\begin{split}
|\psi\rangle&:=  2^{-N/2}\sum_{\underbar{k}} \prod_{\substack{l< j}}^{N}e^{i\phi(k_{l}\wedge k_{j})}|k_1\cdot\cdot\cdot k_N\rangle
\\& =  2^{-N/2}\sum_{\underbar{k}} \prod_{\substack{l< j}}^{N}e^{i\phi\big((k_{l}\oplus 1)\wedge (k_{l}\oplus 1)\big)}|k_1 \oplus 1\cdot\cdot\cdot k_N \oplus 1\rangle.
\end{split}
\end{equation}
% \BA{After some algebra and using the identity in Eq. (\ref{Cp1}) we have 
% \begin{equation}
% \begin{split}
% \langle \psi | \psi_{Y} \rangle &= 2^{-N}(i)^{N}\sum_{k=0}^{N}(-1)^{k}\frac{\binom{N}{k}e^{i \phi \binom{k}{2}}}{e^{i \phi \binom{N-k}{2}}}
% \\& = \bigg(\frac{i}{2}\bigg)^{N}e^{-\frac{i(N-1)N\phi}{2}}\big(1-e^{i(N-1)\phi}\big)^{N}.
% \label{ExpY}
% \end{split}
% \end{equation}}
After some algebra we find the Pauli-$Z$ transformation  
\begin{equation}
\begin{split}
&|\psi_{Z}\rangle=  2^{-N/2}\sum_{\underbar{k}} \prod_{\substack{l< j}}^{N}e^{i\phi(k_{l}\wedge k_{j})}\underbrace{\hat{Z}\otimes \cdot \cdot \cdot \otimes \hat{Z}}_{N}|k_1\cdot\cdot\cdot k_N\rangle
\\& =  2^{-N/2}\sum_{\underbar{k}}(-1)^{\sum_{s}^{N}k_s} \prod_{\substack{l< j}}^{N}e^{i\phi(k_{l}\wedge k_{j})}|k_1\cdot\cdot\cdot k_N \rangle,
\end{split}
\end{equation}
thus yielding 
\begin{equation}
\begin{split}
\langle \psi | \psi_{Z} \rangle &=2^{-N}\sum_{k=0}^{N}(-1)^{k}\sum_{\substack{\underbar{k} \text{ with }\\ k \text{ ones}}}\prod_{l<j}^{N}e^{i \phi \big((k_{l} \wedge k_j)-(k_{l} \wedge k_j)\big)}
\\& = 2^{-N}\sum_{k=0}^{N}(-1)^{k}\binom{N}{k} = 0. 
\label{ExpZ}
\end{split}
\end{equation}

The power series expansion of the exponential matrix, together with the anti-commutation relation of $\hat{Y}$ and $\hat{X}$, yields a simplification of the expectation value in the Heisenberg picture
\begin{equation}
\hspace*{-0.15cm}
\begin{split}
\langle &\hat{A}(\theta) \rangle = \langle \psi | \underbrace{e^{i \frac{\theta}{2}\hat{X}}\hat{Y}e^{-i \frac{\theta}{2}\hat{X}}\otimes \cdot \cdot \cdot \otimes e^{i \frac{\theta}{2}\hat{X}}\hat{Y}e^{-i \frac{\theta}{2}\hat{X}} }_{N}|\psi\rangle
\\& = \langle \psi |\big(\cos \theta \hat{Y}-\sin \theta \hat{Z}\big)\otimes \cdot \cdot \cdot \otimes \big( \cos \theta \hat{Y}-\sin \theta \hat{Z}\big)|\psi\rangle.
\label{ExpV}
\end{split}
\end{equation}

The following is another useful equivalent form of the complete weighted graph state 
\begin{equation}
    \begin{split}
        |\psi\rangle &:=   2^{-N/2}\sum_{\underbar{k}} \prod_{\substack{l< j}}^{N}e^{i\phi k_{l} k_{j}}|k_1\cdot\cdot\cdot k_N\rangle
        \\& = 2^{-N/2}\sum_{\underbar{k}} \text{exp}\bigg(\phi \sum_{l < j}i k_{l}k_{j} \bigg)|k_1\cdot\cdot\cdot k_N\rangle.
    \end{split}
\end{equation}
The following Pauli $Y$-$Z$ transformation on $|\psi\rangle$ leads to
\begin{equation}
\hspace*{-0.2cm}
    \begin{split}
        |\psi_{m}\rangle &:=   2^{-N/2}\sum_{\underbar{k}} \prod_{\substack{l< j}}^{N}e^{i\phi k_{l} k_{j}}(\hat{Y}^{\otimes m}\otimes \hat{Z}^{\otimes N-m})|k_1\cdot\cdot\cdot k_N\rangle
        % \hspace*{-0.45cm}
         \\& ~= 2^{-N/2}i^{m}\sum_{\underbar{k}}(-1)^{\sum_{s}^{N}k_s} \text{exp}\bigg(\phi \sum_{l < j}i k_{l}k_{j} \bigg)
         \\& 
         ~~~~~~\times|k_1\oplus 1\cdot\cdot\cdot k_m\oplus 1 k_{m+1} \cdot \cdot\cdot k_{N}\rangle.
    \end{split}
\end{equation}
Let
\begin{align}
\hspace*{-0.0cm}
q_l := \begin{cases} k_l\oplus1, & 0 \le l \le m, \\ k_l, & m < l \le N,
\end{cases}
\end{align}
this leads to 
\begin{equation}
\hspace*{-0.25cm}
    \begin{split}
    |\psi_m\rangle &=2^{-N/2}i^{m}\sum_{\underbar{k}}(-1)^{\sum\limits_{s=1}^{m}k_s\oplus1 + \sum\limits_{s=m+1}^{N}k_s}   \\&~~~~\times \text{exp}\bigg(\phi \sum_{l < j}i q_{l}q_{j} \bigg)
   |k_1\cdot\cdot\cdot k_m k_{m+1}\cdot\cdot\cdot  k_{N}\rangle,
    \end{split}
\end{equation}
and therefore,
\begin{equation}
\hspace*{-0.25cm}
    \begin{split}
          \langle \psi|\psi_m\rangle &= 2^{-N}i^{m}\sum_{\underbar{k}}(-1)^{\sum\limits_{s=1}^{m}k_s\oplus1 + \sum\limits_{s=m+1}^{N}k_s} \\& \times \text{exp}\bigg(i\phi \sum_{l < j} (q_{l}q_{j}-k_{l}k_{j}) \bigg).
          \label{Exp}
    \end{split}
\end{equation}
Note that for $l+1 \leq j \leq m$, it follows that 
\begin{equation}
    \begin{split}
        q_{l}q_{j}-k_{l}k_{j} &= (k_{l} \oplus 1)(k_{j} \oplus 1)-k_{l}k_{j} \\& = \frac{(-1)^{k_l}+(-1)^{k_j}}{2}.
    \end{split}
\end{equation}
We also make use of the following identity
\begin{equation}
    \begin{split}
        k_{l} \oplus 1 - k_{l}=(-1)^{k_{l}}.
    \end{split}
\end{equation}

Finally, consider the summation in the exponent of (\ref{Exp}),
\begin{multline}
\sum_{l<j}(q_{l}q_{j}-k_{l}k_{j})
       =\sum_{l=1}^{m}\sum_{j=l+1}^{N}(q_{l}q_{j}-k_{l}k_{j})
        \\=\sum_{l=1}^{m}\sum_{j=l+1}^{m}\frac{(-1)^{k_l}+(-1)^{k_j}}{2} \\+\sum_{l=1}^{m}\sum_{j=m+1}^{N}\big((k_l\oplus 1)k_j-k_lk_j \big)
         \\=\sum_{l=1}^{m}\sum_{j=l+1}^{m}\frac{(-1)^{k_l}+(-1)^{k_j}}{2} +\sum_{l=1}^{m}(-1)^{k_l}\sum_{j=m+1}^{N}k_j,    
\end{multline}
therefore Eq.~(\ref{Exp}) now becomes
\begin{multline}
\hspace*{-0.4cm}
\langle \psi|\psi_m\rangle = 2^{-N}i^{m}\sum_{\underbar{k}}(-1)^{\sum\limits_{s=1}^{m}k_s\oplus1 + \sum\limits_{s=m+1}^{N}k_s}\\ \times \text{exp}\Bigg(i\phi\bigg(\frac{1}{2} \sum_{l =1}^{m}\sum_{j=l+1}^{m}\big((-1)^{k_l}+(-1)^{k_{j}}\big)  \\+\sum_{l=1}^{m}(-1)^{k_{l}}\sum_{j=m+1}^{N}k_j\bigg) \Bigg) 
\\= 2^{-N}i^{m}\sum_{a=0}^{N-m}\sum_{b=0}^{m}(-1)^{a+m-b}\binom{N-m}{a}\binom{m}{b}\\ \times \text{exp}\Bigg(i \phi(m-2b)\bigg(\frac{m-1}{2}+a\bigg)\Bigg).
\label{psim3}
\end{multline}
Here $a= \sum_{j=m+1}^{N}k_j$ and $b=\sum_{l=1}^{m}k_{l}$. 
From Eq.~(\ref{ExpV}) it follows that the analytic form of the expectation value (as a function of the unitary parameter $\theta$) reads as
\begin{equation}
\hspace*{-0.1cm}
    \begin{split}
        \langle &\hat{A}(\theta) \rangle = \langle \psi|\sum_{m=0}^{N}\binom{N}{m}(\cos \theta \hat{Y})^{m}(-\sin \theta \hat{Z})^{N-m}|\psi\rangle 
        \\& = \sum_{m=1}^{N}(-1)^{N-m}\binom{N}{m}\cos^{m}\theta\sin^{N-m}\theta\langle \psi| \hat{Y}^{m}\hat{Z}^{N-m}|\psi\rangle.
    \end{split}
\end{equation}
With the last line using Eq.~(\ref{ExpZ}). This simplifies to the form given by Eq.~(\ref{ExpComp}) by substituting in Eq.~(\ref{psim3}) for $\langle \psi| \hat{Y}^{m}\hat{Z}^{N-m}|\psi\rangle$.
\end{proof}

%%%%%%%%%%%%%%%%%%%%%%%%%%%%
%%%%%%%%%%%%%%%%%%%%%%%%%%%%
%%%%%%%%%%%%%%%%%%%%%%%%%%%%
%%%%%%%%%%%%%%%%%%%%%%%%%%%%
% \bibliographystyle{ieeetr}
% \bibliography{bibfile}

\end{document}